\documentclass{rnaastex}

\usepackage[T1]{fontenc}
\usepackage{ae,aecompl}
\usepackage{fontaxes}
\usepackage{graphicx}
\usepackage{amsmath}	
\usepackage{amssymb} 
\newcommand{\HI}{H\,{\sc i}}
\newcommand{\HeI}{He\,{\sc i}}
\newcommand{\HeII}{He\,{\sc ii}}

\newcommand{\SII}{[S\,{\sc ii}]}
\newcommand{\NII}{[N\,{\sc ii}]}
\newcommand{\OI}{[O\,{\sc i}]}
\newcommand{\OII}{[O\,{\sc ii}]}
\newcommand{\OIII}{[O\,{\sc iii}]}
\newcommand{\OVI}{O\,{\sc vi}}

\newcommand{\FeII}{Fe\,{\sc ii}}

\begin{document}

\title{Discovery of a possible symbiotic binary in the Large Magellanic Cloud}

%% Note that the corresponding author command and emails has to come
%% before everything else. Also place all the emails in the \email
%% command instead of using multiple \email calls.
\correspondingauthor{Blesson Mathew}
\email{blesson.mathew@christuniversity.in}

\author{Blesson Mathew}
\affiliation{Department of Physics, Christ University, Bangalore, India}

\author{Warren A.~Reid}
\affiliation{Department of Physics and Astronomy, Macquarie University,\\
  Sydney, NSW 2109, Australia}

\author{R.~E.~Mennickent}
\affiliation{Universidad de Concepci{\'{o}}n, Departamento de Astronom{\'{\i}}a, \\
  Casilla 160-C, Concepci{\'{o}}n, Chile}

\author{D. P. K. Banerjee}
\affiliation{Astronomy and Astrophysics Division, Physical Research Laboratory, \\
Navrangapura, Ahmedabad 380 009, India}

%% Note that RNAAS manuscripts DO NOT have abstracts.

\section{}

We report the discovery of a possible symbiotic star, in the
Large Magellanic Cloud (LMC). The star was detected during the
course of a comprehensive H$\alpha$ survey of the LMC by \citet{Reid12}
aimed at detecting hot emission line stars which resulted in 579
detections, including 469 new discoveries. The H$\alpha$ survey was
followed by multi-object spectroscopy using the 2dF facility at the
Anglo-Australian Telescope (AAT) and FLAMES on the VLT to spectrally classify the objects.
While a variety of emission-line stars were detected, the majority of
them were either Be stars, B[e] stars or  chromospherically active cool
M giants. The object under consideration here, designated as RP 870 (05:23:17.43 -69:38:50.4), was
classified as a B[e] star by \citet{Reid12} on the basis of its spectrum
which displayed several forbidden lines apart from other permitted emission
lines. B[e] stars have all the characteristics of Be stars but they additionally
include forbidden emission lines in their spectra. These could include
\FeII~$\lambda$$\lambda$4244, 4287, 4415, 5273, 7155; \OI~$\lambda$$\lambda$6300,
6363; \NII~$\lambda$$\lambda$5755, 6548, 6584; \SII~$\lambda$$\lambda$4068, 6717,
6731; \OII~$\lambda$$\lambda$7320, 7330 and \OIII~$\lambda$$\lambda$4959, 5007.  

A detailed examination of the red optical spectrum from RP 870 shows
TiO absorption bands that were not picked up by the automated
cross-correlation technique used for the initial object classification.
The spectrum of RP 870 is shown in Figure 1, where apart
from  high ionization emission lines of \HeI, \HeII~and \OIII, one can also
unmistakably see the TiO $\lambda$$\lambda$6180, 7100 molecular absorption
bands associated with a cool component. The presence of such bands
accompanying a hot emission line spectrum is more the  signature of a symbiotic star
rather than that of a B[e] star. \citet{Belczynski00} have laid down the following
criteria for classifying an object as a symbiotic star viz. (i) The presence of
the absorption features of a late-type giant such as TiO, H$_2$O, CO, CN and VO bands
(ii) The presence of strong emission lines of \HI~and \HeI~and emission lines
of ions with an ionization potential of at least 35 eV (e.g. \OIII) and
(iii) The presence of the Raman scattered \OVI~line at 6825 \AA,
even if no features of the cool star (e.g. TiO bands) are found.

The Raman scattered 6825 \AA~line is not seen in the spectrum of RP 870, but it is absent in the spectra
of many symbiotic stars \citep[see the $\sim$200 spectra shown in][]{Munari02}.
The collective  signatures of a hot component
(high excitation/ionization lines) and of a cool component (TiO molecular bands)
are seen in RP 870, from which we propose it as a symbiotic star.
Since known symbiotic systems are rare in the LMC, possibly
less than a dozen are known \citep{Miszalski14b,Hajduk15,Angeloni09},
we thought the present  detection to be interesting enough to be reported.

Figure 1 also shows a deep H$\alpha$ image of the field around RP 870
obtained by \citet{Reid12} which shows the strong H$\alpha$ emission
from RP 870. We used this image, and also inspected other archival images
like 2MASS etc, to check whether the spectrum of RP 870 was contaminated with
the light of any other nearby source. It was prudent to ensure this since the
tip of each optical fiber of 2dF sees 2 arcsec of the sky.
However, we do not find any reason to suspect source contamination. 

\begin{figure}[hb]
\begin{center}
\includegraphics[width=8.5cm,height=7.5cm]{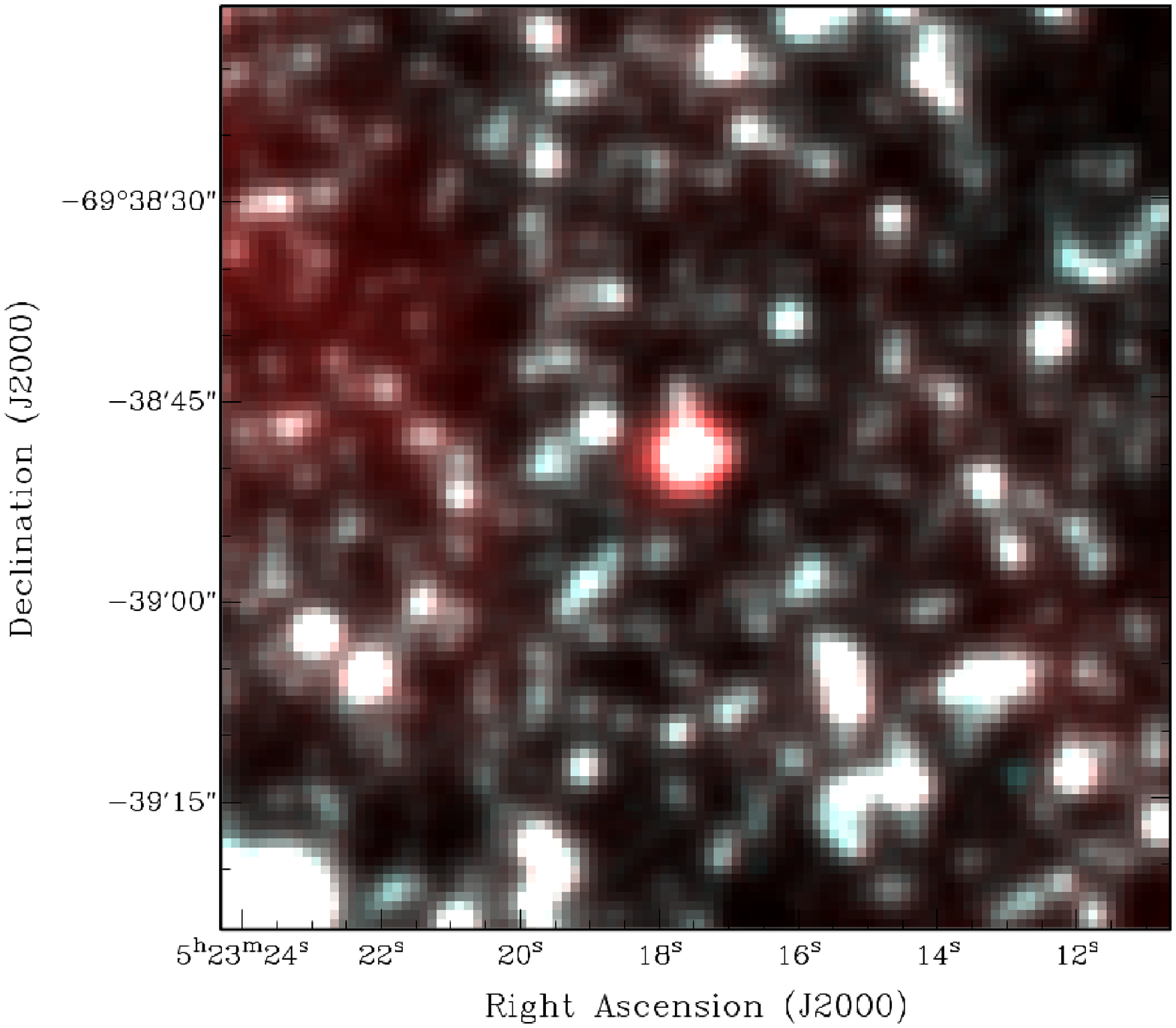}
\includegraphics[width=8cm,height=8cm]{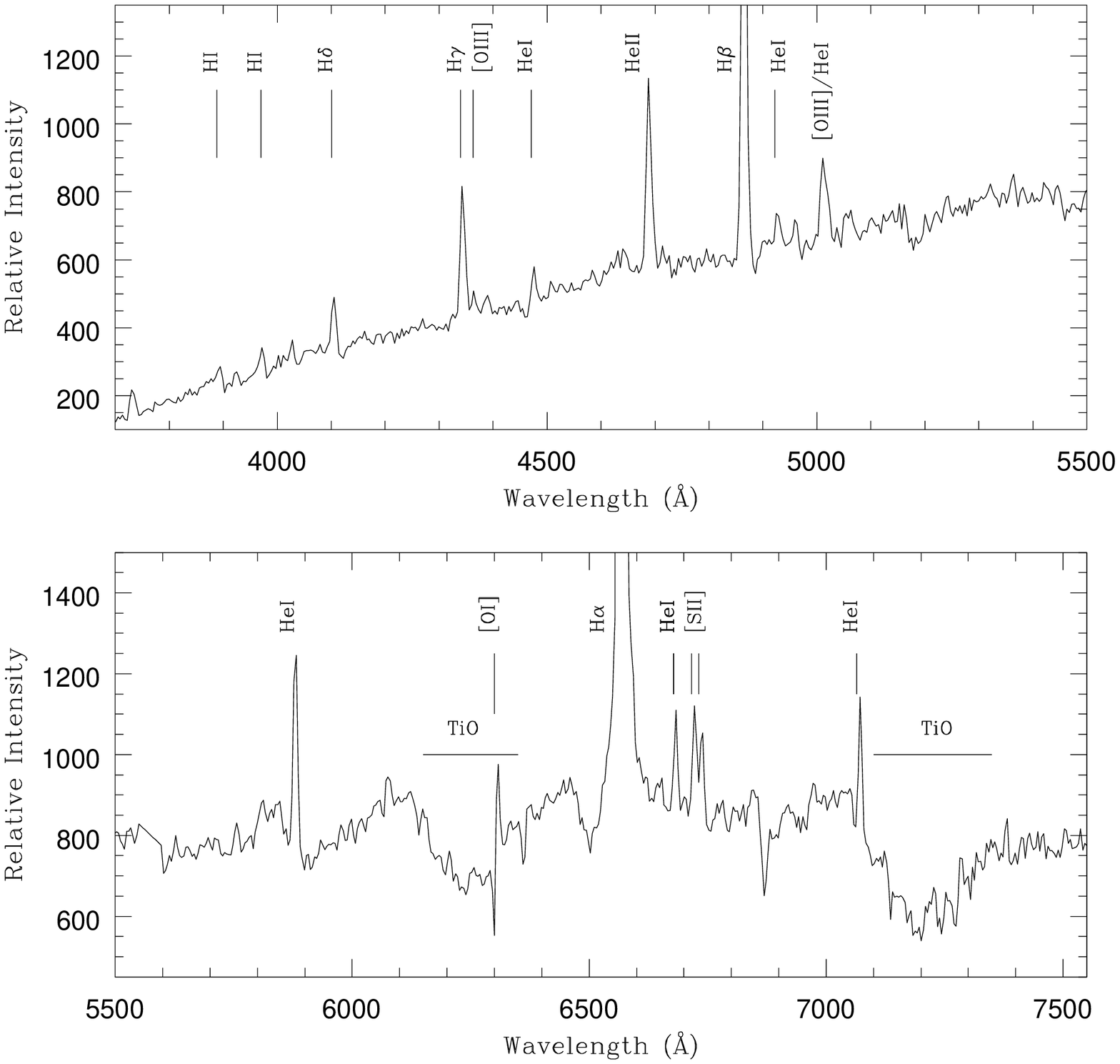}
\caption{Left: Left: The combined H$\alpha$ + Short Red (SR) broadband
  image of RP~870 from the UKST deep-stacked maps is shown.
  The object was imaged in narrow-band H$\alpha$ and broad-band
  Short-Red with United Kingdom Schmidt Telescope (UKST).
  The twelve highest quality and well-matched UK Schmidt Telescope
  2-hour H$\alpha$ exposures and six 15-minute equivalent SR-band
  exposures were selected and overlayed with false colors in
  order to detect emission objects \citep[see][for more details]{Reid06}.
  H$\alpha$ has been assigned a red color while the SR is both blue and
  green, resulting in a white color for continuum sources.
  The H$\alpha$ excess from RP 870 (at the center of the image) can be seen
  from the redness it displays.  Right: AAT 2dF spectrum of RP~870.\label{fig:1}}
\end{center}
\end{figure}

\acknowledgments

The authors wish to thank the Australian Astronomical Observatory
for observing time on the AAT.
R.E.M. acknowledges support by VRID-Enlace 216.016.002-1.0 and the
BASAL Centro de Astrof{\'{i}}sica y Tecnolog{\'{i}}as Afines (CATA) PFB--06/2007.

\end{document}